\documentstyle[amsfonts,12pt]{article}
\begin{document}
\title{Conditionally exactly solvable potentials:\\ A supersymmetric
  construction method} 
\author{
Georg Junker\\
Institut f\"ur Theoretische Physik, 
Universit\"at Erlangen-N\"urnberg,\\ 
Staudtstr.\ 7, D-91058 Erlangen, Germany.\\~\\ 
Pinaki Roy\\[2mm]
Physics and Applied Mathematics Unit,\\
Indian Statistical Institute, Calcutta 700035, India.
}
\maketitle
%%%%%%%%%%%%%%%%%%%%%%%%%%%%%%%%%%%%%%%%%%%%%%%%%%%%%%%%%%%%%%%%%%%%%%%%%%%%%%
\begin{abstract}
We present in this paper a rather general method for the construction of
so-called conditionally exactly solvable potentials. This method is
based on algebraic tools known from supersymmetric quantum mechanics. Various
families of one-dimensional potentials are constructed whose corresponding
Schr\"odinger eigenvalue problem can be solved exactly under certain
conditions of the potential parameters. Examples of quantum systems on the
real line, the half line as well as on some finite interval are studied in
detail. 
\end{abstract}
%%%%%%%%%%%%%%%%%%%%%%%%%%%%%%%%%%%%%%%%%%%%%%%%%%%%%%%%%%%%%%%%%%%%%%%%%%%%%%
\section{Introduction}
Since the advent of quantum mechanics there has been interest in
quantum models whose corresponding Schr\"odinger equation can be solved
exactly. To be more precise, by exactly solvable we mean that the 
spectral properties, that is, the eigenvalues and eigenfunctions, of the 
Hamiltonian characterizing the quantum system under consideration can be
given in an explicit and closed form. The most important examples are the
harmonic oscillator and the hydrogen atom. An first attempt in finding such
systems has been initiated by Schr\"odinger \cite{Schr41} 
himself and is now know as the factorization method \cite{InHu51}. 
This factorization method has been revived during the
last two decades in connection with supersymmetric quantum mechanics
\cite{Ju96}.  
In particular, the factorization condition which is a condition on the 
quantum mechanical potential for its exact solvability has been rediscovered
and is now known as the so-called shape-invariance condition \cite{Gen83}. 
In fact, there have been several attempts in finding additional
shape-invariant potentials besides those already given by Infeld and
Hull \cite{InHu51}. 

Other methods which are also closely related to supersymmetric (SUSY) quantum
mechanics are based on the idea of finding pairs of (essentially) isospectral
Hamiltonians \cite{AbMo80,Mie84,Nie84,LuPu86, Roy86}.  
One of these methods, the Darboux method, is based on the existence of an
operator $A$ and its adjoint $A^\dagger$ which act as transformation operators
between a pair of self-adjoint Hamiltonians $H_\pm$ \cite{Dar1882,Dei78}:
\begin{equation}
AH_- = H_+A\;,\quad H_-A^\dagger=A^\dagger H_+\;.
\label{1.1}
\end{equation}
Obviously, $H_+$ and $H_-$ are essentially isospectral, that is, there spectra
coincide except of a possible additional vanishing eigenvalue. Knowing, for 
example the eigenfunctions of $H_+$ one can immediately obtain those of $H_-$
with the help of the transformation operator $A^\dagger$. This Darboux method,
which has originally been applied with linear first-order differential 
operators $A$, has recently been extended to higher-order differential 
operators where it is called $N$-extended Darboux transformation (with $N$ 
standing for the highest order of the momentum operator appearing in $A$)
\cite{AnIoSp93,BaSa95}. 

Another different method for constructing  exactly solvable systems has been
suggested by Abraham and Moses \cite{AbMo80} and is based on the inverse
method. As in the 
Darboux method one starts with a given exactly solvable Hamiltonian and 
constructs a new one whose spectral properties follow from those of the 
starting Hamiltonian. Applying this approach to SUSY quantum systems
it is equivalent to the Darboux method \cite{Nie84}.

In this paper we develop yet another method for constructing so-called
conditionally exactly solvable systems \cite{SoDu83}. This method, which is
based on 
the SUSY formulation of one-dimensional quantum systems has recently been
suggested by us in \cite{JuRo97}. It is the aim of this paper to present the
detailed 
ideas of this approach and to apply it to various physically relevant model
systems on the real line, the half line and those on a finite interval.
In particular, we will show that many of the newly found exactly solvable
potentials contain as special cases also those found by the other two methods
mentioned above.
 
In the next section we will briefly review the basic algebraic tools of SUSY
quantum mechanics \cite{Ju96}, which will be used in the general construction
method presented in Section 3. The remaining three sections present a detailed 
discussion of examples on the real line, the half line and finite intervals.
To be more explicit, in Section 4 we construct the most general class (within 
our approach) of SUSY partner potentials for the linear harmonic,
the Morse and the symmetric Rosen-Morse oscillator. Section 5 contains the
corresponding results for the 
radial harmonic oscillator and the radial Coulomb problem. In Section 6 we
consider the symmetric P\"oschl-Teller oscillator as an example on the finite
interval $[-\pi/2,\pi/2]$.  

%%%%%%%%%%%%%%%%%%%%%%%%%%%%%%%%%%%%%%%%%%%%%%%%%%%%%%%%%%%%%%%%%%%%%%%%%%%%%%

\section{Supersymmetric quantum mechanics}
\setcounter{equation}{0}
In this section we briefly review the basic concepts of Witten's model of 
supersymmetric quantum mechanics \cite{Wit81,Ju96}. This model consists of a
pair of standard Schr\"odinger Hamiltonians
\begin{equation}
H_\pm=-\frac{1}{2}\,\frac{d^2}{dx^2}+V_\pm(x)
\label{2.1}
\end{equation}
which act on the Hilbert space ${\cal H}$ of square integrable functions 
over a given configuration space. In this paper we will consider systems
on the real line ${\mathbb R}$, on the positive half line ${\mathbb R}^+$, and
on the finite interval $x\in [-\frac{\pi}{2},\frac{\pi}{2}]$. 
In the latter two cases we will impose
Dirichlet boundary conditions, that is, the Hilbert spaces are given by
${\cal H}=L^2({\mathbb R})$, 
${\cal H}=\{\psi\in L^2({\mathbb R}^+)|\psi(0)=0\}$, and 
${\cal H}=\{\psi\in
L^2([-\frac{\pi}{2},\frac{\pi}{2}])|\psi(\pm\frac{\pi}{2})=0\}$,
respectively. 
The so-called SUSY partner potentials in (\ref{2.1})
are expressed in terms of the real-valued SUSY potential $W$ and its 
derivative $W'=dW/dx$,
\begin{equation}
V_\pm(x)= \frac{1}{2}\Bigl(W^2(x)\pm W'(x)\Bigr)\;.
\label{2.2}
\end{equation}
Introducing the supercharge operators
\begin{equation}
A=\frac{1}{\sqrt{2}}\left(\frac{d}{dx}+W(x)\right)\;,\quad
A^\dagger=\frac{1}{\sqrt{2}}\left(-\frac{d}{dx}+W(x)\right)
\label{2.3}
\end{equation}
the SUSY partner Hamiltonians factorize as follows
\begin{equation}
H_+=AA^\dagger\geq 0\;,\quad H_-=A^\dagger A\geq 0 
\label{2.4}\end{equation}
and obviously obey the relation (\ref{1.1}). As a consequence
$H_+$ and $H_-$ are essentially isospectral, that is, their strictly
positive energy eigenvalues coincide. In addition one of the two Hamiltonians
may have a vanishing eigenvalue. In this case, SUSY is said to be unbroken
and by convention \cite{Ju96} (via an appropriate choice of an overall sign in
$W$) this ground state then belongs to $H_-$. This convention implies that
$\exp\left\{\int dx\, W(x)\right\} \notin {\cal H}$.

Let us be more explicit and denote the eigenfunctions and eigenvalues of 
$H_\pm$ by $\psi^\pm_n$ and $E^\pm_n$, respectively. That is,
\begin{equation}
H_\pm\psi^\pm_n(x)=E^\pm_n\psi^\pm_n(x)\;,\quad n=0,1,2,\ldots\;.
\label{2.5}
\end{equation}
For simplicity we consider only the discrete part of the spectrum here.
However, relations similar to those given below are also valid for the 
continuous part.
In the case of unbroken SUSY (within the aforementioned convention) 
the zero-energy eigenstate of the SUSY system belongs to $H_-$ and the
corresponding ground state has the properties
\begin{equation}
E_0^-=0\;,\quad \psi_0^-(x)=C\exp\left\{-\int dx\, W(x)\right\}\in {\cal H}
\label{2.6}
\end{equation}
with $C$ denoting the normalization constant. 
The remaining spectrum of $H_-$
coincides with the complete spectrum of $H_+$ and the corresponding 
eigenfunctions are related by SUSY transformations which are generated by the
supercharge operators (\ref{2.3}):
\begin{equation}
E_{n+1}^-=E_n^+>0\;, \quad
\psi_{n+1}^-(x)=(E_n^+)^{-1/2}A^\dagger\psi_n^+(x)\;,\quad
\psi_{n}^+(x)=(E_{n+1}^-)^{-1/2}A\psi_{n+1}^-(x)\;.
\label{2.7}
\end{equation}
In the case of broken SUSY $H_+$ and $H_-$ are strictly isospectral and the 
eigenfunctions are also related by SUSY transformations:
\begin{equation}
E_{n}^-=E_n^+>0\;,\quad 
\psi_{n}^-(x)=(E_n^+)^{-1/2}A^\dagger\psi_n^+(x)\;,\quad
\psi_{n}^+(x)=(E_{n}^-)^{-1/2}A\psi_{n}^-(x)\;.
\label{2.8}
\end{equation}

With the help of the relations (\ref{2.6}) and (\ref{2.7}) or 
(\ref{2.8}) it is obvious that knowing the spectral properties of, say 
$H_+$, one immediately obtains the complete spectral properties of the SUSY 
partner Hamiltonian $H_-$. These facts will be our basis for the
general construction method of conditionally exactly solvable 
potentials, by which we mean that the eigenvalues and eigenfunctions
of the corresponding Schr\"odinger Hamilton\-ian can be given in an explicit
closed form (under certain conditions obeyed by the potential parameters). 

%%%%%%%%%%%%%%%%%%%%%%%%%%%%%%%%%%%%%%%%%%%%%%%%%%%%%%%%%%%%%%%%%%%%%%%%%%%%%%

\section{The construction method}
\setcounter{equation}{0}
In this section we present a rather general method for the construction of
conditionally exactly solvable potentials using the SUSY transformations 
between the eigenstates of the SUSY partner Hamiltonians $H_\pm$. 
The basic idea is as follows. Let us look for some SUSY potential $W$ such that
under certain conditions on its parameters the corresponding partner potential
$V_+$ becomes an exactly solvable one. For example, one of the shape-invariant
potentials known form the factorization method \cite{InHu51,Ju96}. As a
consequence the spectral 
properties of the associated Hamilton\-ian $H_+$ are known exactly. 
From the given SUSY potential $W$ also follows the corresponding partner
potential $V_-$ and its associate Hamiltonian $H_-$. As we will see below,
this potential is in general not shape-invariant but still exactly solvable
via the SUSY transformations (\ref{2.7}) or (\ref{2.8}). 

In order to find an appropriate class of SUSY potentials we make the ansatz
\begin{equation}
W(x)=\Phi(x)+f(x)
\label{3.1}
\end{equation}
where $\Phi$ is chosen such that  for 
$f\equiv 0$ the corresponding partner potentials $V_\pm$ belong to the known 
class of shape-invariant exactly solvable ones. 
For a non-vanishing $f$ we have
\begin{equation}
\textstyle
V_+(x)=\frac{1}{2}\Bigl[\Phi^2(x)+\Phi'(x)+f^2(x)+2\Phi(x)f(x)+f'(x)\Bigr]\;.
\label{3.2}
\end{equation}
If we now impose on $f$ the condition that it obeys the following generalized 
Riccati equation
\begin{equation}
f^2(x)+2\Phi(x)f(x)+f'(x)=b\;,
\label{3.3}
\end{equation}
where, for the moment, $b$ is assumed to be an arbitrary real constant, then
the two partner potentials take the form
\begin{eqnarray}
V_+(x)&=&\textstyle\frac{1}{2}\Bigl[\Phi^2(x)+\Phi'(x)+b\Bigr]\;,
\label{3.4}\\
V_-(x)&=&\textstyle\frac{1}{2}\Bigl[\Phi^2(x)-\Phi'(x)-2f'(x)+b\Bigr]\;.
\label{3.5}
\end{eqnarray}
Obviously, $V_+$ is, up to the additive constant $b/2$, a shape-invariant
potential and therefore exactly solvable. With the help of the SUSY 
transformation we can now also solve the eigenvalue problem for $H_-$ 
for the above given potential $V_-$ which, due to the additional $x$-dependent
term $f'$ will in general be a new non-shape-invariant potential.
At this step we already realize that the free parameter $b$ has to be bounded
below, as SUSY already requires a strictly positive Hamiltonian $H_+$. This is
a first condition on a parameter contained in $V_-$ and already justifies to
call it a conditionally exactly solvable (CES) potential.

The crucial problem in finding new CES potentials is to find the most general
solution of the  generalized Riccati equation (\ref{3.3}). For this reason we
linearize this equation by making the ansatz
\begin{equation}
  \label{3.6}
  f(x)=\frac{d}{dx}\,\log u(x)=\frac{u'(x)}{u(x)}\;.
\end{equation}
which brings it into the form of a homogeneous linear
second-order differential equation
\begin{equation}
  \label{3.7}
  u''(x)+2\,\Phi(x)\,u'(x)-b\,u(x)=0\;.
\end{equation}
The general solution of this equation is given by a linear combination
of two linearly independent fundamental solutions
\begin{equation}
  \label{3.8}
  u(x)=\alpha\, u_1(x)+\beta\, u_2(x)\;.
\end{equation}
Hence, besides the parameters contained in $\Phi$ and the parameter $b$ the
new CES potential $V_-$ will also depend on the real parameters $\alpha$ and
$\beta$. Note however, that only the quotient $\alpha/\beta$ or $\beta/\alpha$
will enter $V_-$ as a relevant parameter. In other words, depending on the 
actual situation one of these two parameters can be chosen (without loss of
generality) to unity. The remaining parameters, however, will in general be
not arbitrary real numbers and have to be chosen such that the
corresponding supercharges
\begin{equation}
  \label{3.9}
  A=\frac{1}{\sqrt{2}}\left(\frac{d}{dx}+\Phi(x)+\frac{u'(x)}{u(x)}\right)\;,
  \quad
  A^\dagger=\frac{1}{\sqrt{2}}\left(-\frac{d}{dx}+\Phi(x)+\frac{u'(x)}{u(x)}
  \right)
\end{equation}
are well-defined operators leaving the Hilbert space invariant,
$A:{\cal H}\to{\cal H}$, $A^\dagger:{\cal H}\to{\cal H}$.
A sufficient condition for that is to allow only for nonvanishing solutions
(\ref{3.8}). Thus the parameters have to be chosen such that $u$ is (without
loss of generality) a strictly positive function. Indeed, this condition also
guarantees us that the potential
\begin{equation}
  \label{3.10}
  V_-(x)=\frac{1}{2}\,\Phi^2(x)-\frac{1}{2}\,\Phi'(x)+\frac{u'(x)}{u(x)}\left(
2\,\Phi(x)+\frac{u'(x)}{u(x)}\right) -\frac{b}{2}
\end{equation}
does not have singularities inside the configuration 
space. So $H_+$ and $H_-$ have indeed a common domain ${\cal H}$. This
condition is actually the most difficult part in our approach. 

For all shape-invariant SUSY potentials, which we have considered,
eq.(\ref{3.7}) can 
be reduced to a hypergeometric or confluent hypergeometric differential
equation. That is, the two fundamental solutions $u_1$ and $u_2$ in
(\ref{3.8}) are expressed in terms of hypergeometric or confluent
hypergeometric functions. Finding the proper linear combination leading
to a strictly positive solution is very difficult and in general can be
obtained only by inspection (numerically and/or via the asymptotic behaviour
at the boundaries of the configuration space). 

Besides the above mentioned necessary conditions on the potential parameters
$b$, $\alpha$, $\beta$ and possible additional ones contained in $\Phi$, we
will 
further restrict these parameters in the following respect. Let us assume that
the SUSY quantum system (\ref{2.1}) is unbroken (broken) for $f=0$. Then we
consider only those values of the parameters for which the system with 
$f\neq 0$ remains to have unbroken (broken) SUSY. Hence, due to our
ground-state convention, we have the following additional conditions:
 \begin{equation}
      \label{3.11}
      \begin{array}{l}
\displaystyle
\exp\left\{\int dx\, W(x)\right\}=u(x)\exp\left\{\int dx\,
\Phi(x)\right\}\notin {\cal H}\quad
\mbox{for broken and unbroken SUSY}\;,\\[4mm]
\displaystyle
\exp\left\{-\int dx\, W(x)\right\}=[u(x)]^{-1}
\exp\left\{-\int dx\,\Phi(x)\right\}\notin {\cal H}\quad
\mbox{for broken SUSY}\;,\\[4mm]
\displaystyle
\exp\left\{-\int dx\, W(x)\right\}=[u(x)]^{-1}
\exp\left\{-\int dx\,\Phi(x)\right\}\in {\cal H}\quad
\mbox{for unbroken SUSY}\;.
      \end{array}
    \end{equation}

In the following we will consider several examples on the real line, the
positive half line and a finite interval. Both, unbroken as well as broken
SUSY systems will be discussed.

%%%%%%%%%%%%%%%%%%%%%%%%%%%%%%%%%%%%%%%%%%%%%%%%%%%%%%%%%%%%%%%%%%%%%%%%%%%%%%

\section{Quantum systems on the real line}
\setcounter{equation}{0}
In this section we will consider two examples on the real line in some
detail. These are the linear and the Morse oscillator, which both have a
unbroken SUSY. Note that there are no known shape-invariant potentials on
${\mathbb R}$ which allow for a broken SUSY. Finally, we also briefly
summarize some results for the symmetric Rosen-Morse oscillator.

%%%%%%%%%%%%%%%%%%%%%%%%%%%%%%%%%%%%%%%%%%%%%%%%%%%%%%%%%%%%%%%%%%%%%%%%%%%%%%

\subsection{The linear harmonic oscillator}
The first SUSY system we are considering is characterized by a linear SUSY 
potential $\Phi(x)=x$ which gives rise to a unbroken SUSY with potential
\begin{equation}
V_+(x)=\textstyle\frac{1}{2}\left(x^2+b+1\right)\;.
\label{4.1}
\end{equation}
The energy eigenvalues and eigenfunctions of the corresponding Hamiltonian read
\begin{equation}
E_n^+=n+b/2 +1\;,\quad
\psi_n^+(x)=\left[\sqrt{\pi}\,2^nn!\right]^{-1/2}H_n(x)\exp\{-x^2/2\}\;,
\label{4.2}
\end{equation}
where $H_n$ denotes the Hermite polynomial of order $n\in{\mathbb N}_0$. 
Clearly, positivity of $H_+$ implies the condition $b>-2$.

The general solution of (\ref{3.7}) can be given in terms of confluent
hypergeometric functions \cite{MaObSo66},
\begin{equation}
\textstyle
u(x)=\alpha\,_1F_1\left(-\frac{b}{4},\frac{1}{2},-x^2\right) + 
\beta\,x\,_1F_1\left(\frac{2-b}{4},\frac{3}{2},-x^2\right)\;,
\label{4.3}
\end{equation}
and has the following asymptotic behaviour for $x\to\pm\infty$
\begin{equation}
  \label{4.3a}
  u(x)=|x|^{b/2}\left(\alpha\,\frac{\Gamma(1/2)}{\Gamma(\frac{b+2}{4})}+
       \beta\,\frac{\Gamma(3/2)}{\Gamma(\frac{b}{4}+1)}\right)(1+O(|x|^{-1})\;.
\end{equation}
Here and in the following $\Gamma$ denotes Euler's gamma function.
From this asymptotic behaviour the condition on the parameters $\alpha$ and
$\beta$ for a strictly non-vanishing $u$ reads
$|\beta/\alpha|<2\,\Gamma(\frac{b}{4}+1)/\Gamma(\frac{b+2}{4})$. Note that the
right-hand side of this inequality is positive as $b>-2$ and that $\alpha$
must not vanish, that is, it can be chosen equal to unity, $\alpha=1$.

The potential $V_-$ can be obtained from (\ref{3.10}) and explicitly reads
\begin{equation}
  \label{4.3b}
  V_-(x)=\frac{1}{2}\, x^2 -\frac{b+1}{2}+\frac{u'(x)}{u(x)}\left[2\, x +
  \frac{u'(x)}{u(x)} \right]
\end{equation}
where $u$ is given in (\ref{4.3}). The eigenvalues and
eigenfunction for the associated partner Hamiltonian $H_-$ are found via
(\ref{2.6}) and (\ref{2.7}) as SUSY remains unbroken:
\begin{equation}
\begin{array}{ll}
E_0^-=0\;,\quad & \displaystyle 
\psi_0^-(x)=\frac{C}{u(x)}\exp\{-x^2/2\}\;,\\[2mm]
E_{n+1}^-=E_n^+ \;,\quad &  \displaystyle 
\psi_{n+1}^-(x)=
\frac{\exp\{-x^2/2\}}{\left[\sqrt{\pi}\,2^{n+1}n!(n+b/2 + 1)\right]^{1/2}}
\left(H_{n+1}(x)+H_n(x)\frac{u'(x)}{u(x)}\right).
\end{array}
\label{4.4}
\end{equation}
Figure 1 presents a graph of this family of potentials for 
$b\in[-2.5,3]$, $\alpha=1$ and
$\beta \equiv \beta(b)
= 1.5\times\Gamma(\frac{b}{4}+1)/\Gamma(\frac{b+2}{4})$.
It clearly shows singularities for $b\leq -2$ as expected. 
In Figure 2 we keep 
$b=-1.9$ fixed and display the potential $V_-$ for various values of the
asymmetry parameter $\beta$. Again singularities appear for $|\beta|\geq
2\Gamma(\frac{b}{4}+1)/\Gamma(\frac{b+2}{4})\simeq 0.08569$. 
Let us note here that
the potential (\ref{4.3b}) has previously been considered by Hongler and Zheng
\cite{HoZe82} in connection 
with an exactly solvable Fokker-Planck problem, which is closely related to
Witten's SUSY quantum mechanics \cite{Ju96}. 

Special cases of $V_-$ have also previously been found with the methods
mentioned in the introduction. For example, the special values $b=0$, 
$\alpha=\gamma$ and $\beta=1$ lead to $u(x)=\gamma+(\sqrt{\pi}/2)\mbox{Erf}(x)$
(Erf denotes the error function) which is the result of Mielnik
\cite{Mie84}. For $b=4N$, $N\in{\mathbb N}$, $\alpha=1$ and $\beta=0$ the
conditionally exactly solvable potential reads
\begin{equation}
  \label{4.5}
V_-(x)=\frac{x^2}{2}+8N(2N-1)\,\frac{H_{2N-2}(ix)}{H_{2N}(ix)}-
16N^2\left(\frac{H_{2N-1}(ix)}{H_{2N}(ix)}\right)^2 +2N-\frac{1}{2}
\end{equation}
which has previously been obtained by Bagrov and Samsonov \cite{BaSa95} via
the $N$-order Darboux method. See also \cite{JuRo97} where, in particular, the
cases $N=1$ and 2 have been discussed.

%%%%%%%%%%%%%%%%%%%%%%%%%%%%%%%%%%%%%%%%%%%%%%%%%%%%%%%%%%%%%%%%%%%%%%%%%%%%%%

\subsection{The Morse oscillator}
As a second example we consider the Morse oscillator which is
characterized by the SUSY potential
\begin{equation}
  \label{4.6}
  \Phi(x)=\gamma - e^{-x}\;,\quad \gamma > 0\;,
\end{equation}
where the condition on the parameter $\gamma$ results from our ground-state
convention (see Section 2). Changing from parameter $b$ to
\begin{equation}
  \label{4.7}
  \rho = \sqrt{\gamma^2 + b}
\end{equation}
the corresponding potential (\ref{3.4}) reads
\begin{equation}
  \label{4.8}
  V_+(x)=\frac{1}{2}\left(e^{-2x}-(2\gamma-1)e^{-x}+\rho^2\right).
\end{equation}
The (discrete) spectral properties of the associated Hamiltonian $H_+$ are
\begin{equation}
  \label{4.9}
  \begin{array}{l}
    E_n^+=-\frac{1}{2}\left(\gamma-n-1\right)^2+\frac{\rho^2}{2}\;,\quad
    n=0,1,2,\ldots <\gamma -1\;,\\[2mm]
    \displaystyle
    \psi_n^+(x)=
    \left[\frac{(2\gamma-2n-2)\Gamma(n+1)}{\Gamma(2\gamma-n-1)}\right]^{1/2} 
    2^{\gamma-n-1}\exp\{-e^{-x}-x(\gamma-n-1)\}
    L_n^{(2\gamma-2n-2)}(2e^{-x})\;,
  \end{array}
\end{equation}
with $L_n^{\nu}$ denoting the generalized Laguerre polynomial of order $n$
\cite{MaObSo66}. Obviously, positivity of $H_+$ implies the condition 
\begin{equation}
  \label{4.10}
  \rho > \gamma -1\;.
\end{equation}
With the above SUSY potential (\ref{4.6}) the differential equation
(\ref{3.7}) can be reduced to that of the confluent hypergeometric equation
and in turn the general solution reads
\begin{equation}
  \label{4.11}
  u(x)=\alpha e^{-x(\gamma+\rho)}\,_1F_1(\gamma+\rho,1+2\rho,-2e^{-x}) +
       \beta  e^{-x(\gamma-\rho)}\,_1F_1(\gamma-\rho,1-2\rho,-2e^{-x})\;,
\end{equation}
which has the following asymptotic behaviour for $x\to -\infty$:
\begin{equation}
  \label{4.13}
  u(x)=\alpha\,\frac{\Gamma(1+2\rho)}{2^{\gamma+\rho}\Gamma(1-\gamma+\rho)} 
+ \beta\,\frac{\Gamma(1-2\rho)}{2^{\gamma-\rho}\Gamma(1-\gamma-\rho)} +
O(e^x)\;.
\end{equation}
From the asymptotic behaviour of $u$ for $x\to+\infty$, which can
trivially be extracted from (\ref{4.11}), and the form of the SUSY ground-state
wavefunction 
\begin{equation}
  \label{4.14}
  \psi_0^-(x)=\frac{C}{u(x)}\,\exp\{-\gamma x - e^{-x}\}
\end{equation}
one finds that SUSY remains unbroken iff $\beta\neq 0$. Hence, we can set it
equal to unity, $\beta =1$. 
The positivity condition of $u$ can, with the help of the
relation (\ref{4.13}), be translated into conditions on the remaining
parameters. These are
\begin{equation}
  \label{4.12}
 \rho>\gamma-1\;,\quad \frac{\Gamma(1-2\rho)}{\Gamma(1-\rho-\gamma)} >
 0\;,\quad
 \alpha > - 2^{2\rho} \,
\frac{\Gamma(1-2\rho)\Gamma(1+\rho-\gamma)}
     {\Gamma(1+2\rho)\Gamma(1-\rho-\gamma)}\;,
\end{equation}
which have to be obeyed simultaneously.

In Figure 3 we have shown the family of potentials 
\begin{equation}
  V_-(x)=\frac{1}{2}\,e^{-2x}-\left(\gamma + \frac{1}{2}\right)e^{-x} +
  \gamma^2 - \frac{\rho^2}{2} 
  + \frac{u'(x)}{u(x)}\left(2\gamma - 2e^{-x}+\frac{u'(x)}{u(x)}\right)
\end{equation}
for $\alpha=0$, $\gamma=1$ and $\rho\in [0,4]$.  
Note that from (\ref{4.12}) the allowed values of $\rho$ for
the given $\alpha$ and $\gamma$ are
$\rho\in\cup_{k=0}^\infty]2k+\frac{1}{2}, 2k+\frac{3}{2}[$. These admissible
ranges of $\rho$ are clearly visible in Figure 3.
Figure 4 shows the graph of $V_-$ for the cases $\alpha\neq 0$ and
$\gamma=\rho=3$. Note that the last condition in (\ref{4.12}) now explicitly
reads $\alpha>-4/45 =-0.08889$.
The violation of this condition is also clearly visible in Figure 4 via the
singularities in $V_-$.

To complete the discussion of this example we finally give the discrete
spectral 
properties of the corresponding partner Hamiltonian $H_-$. As SUSY remains
unbroken the ground-state energy vanishes, $E_0^-=0$, and the corresponding
eigenfunction is given in (\ref{4.14}). For the excited states the discrete
spectrum is given by $E_{n+1}^-=E_n^+$ and the associated wavefunctions
explicitly read 
\begin{equation}
\label{4.17}
  \begin{array}{ll}
\psi_{n+1}^-(x)=&\displaystyle
\left[\frac{(2\gamma-2n-2)\Gamma(n+1)}{(\rho^2-(\gamma-n-1)^2)
\Gamma(2\gamma-n-1)}\right]^{1/2} 2^{\gamma-n-1}\exp\{-e^{-x}-x(\gamma-n-1)\}
\\[4mm]
&\displaystyle\times
\left[(n+1)L_{n+1}^{(2\gamma-2n-2)}(2e^{-x})+
\frac{u'(x)}{u(x)}\,L_{n}^{(2\gamma-2n-2)}(2e^{-x})\right]\;.
  \end{array}
\end{equation}

%%%%%%%%%%%%%%%%%%%%%%%%%%%%%%%%%%%%%%%%%%%%%%%%%%%%%%%%%%%%%%%%%%%%%%%%%%%%%%

\subsection{The symmetric Rosen-Morse oscillator}
As a last example on the real line let us briefly discuss the symmetric
Rosen-Morse potential (sometimes also called modified P\"oschl-Teller
potential) which is characterized be the SUSY potential
\begin{equation}
  \label{4.18}
  \Phi(x)= \gamma \tanh (x)\;,\quad \gamma > 0\;.
\end{equation}
The corresponding potential $V_+$ reads
\begin{equation}
V_+(x)=-\frac{\gamma (\gamma -1)}{2\cosh^2 x} + \frac{\gamma^2+b}{2}
  \label{4.19}
\end{equation}
and for $\gamma\in{\mathbb N}$ is known to belong to the class of
reflectionless potentials, which are, for example, important for the
construction of explicit solutions of the Korteweg-deVries equation
\cite{Dra83}.  

For the above SUSY potential (\ref{3.7}) can be reduced to
Legendre's differential equation and the general solution is given by
\begin{equation}
  \label{4.20}
  u(x)=\cosh^{-\gamma}(x)
\left[\alpha P_{\gamma-1}^{(\gamma^2 + b)^{1/2}}(\tanh x) 
     + \beta Q_{\gamma-1}^{(\gamma^2 + b)^{1/2}}(\tanh x)\right]
\end{equation}
where $P_\nu^\mu$ and $Q_\nu^\mu$ denote Legendre functions as defined in
\cite{MaObSo66}. 
We leave it to the reader to investigate the proper admissible ranges
for the potential parameters $b,\alpha,\beta$ and $\gamma$, and only remark
that the family of partner potentials
\begin{equation}
  V_-(x)=-\frac{\gamma(\gamma + 1)}{2\cosh^2x}+\frac{\gamma^2 - b}{2}+
\frac{u'(x)}{u(x)}\left(2\gamma\tanh x + \frac{u'(x)}{u(x)}\right)
\end{equation}
will contain new reflectionless potentials (via the choice $\gamma\in{\mathbb
N}$) and thus may, for example, allow to find new explicit solutions for the
Korteweg-deVries equation. 

%%%%%%%%%%%%%%%%%%%%%%%%%%%%%%%%%%%%%%%%%%%%%%%%%%%%%%%%%%%%%%%%%%%%%%%%%%%%%%

\section{Quantum systems on the positive half line}
\setcounter{equation}{0}
As examples of new CES potentials on the positive half line we consider in
this section the radial harmonic oscillator, which allows for unbroken as well
as broken SUSY, and the radial hydrogen atom problem.

%%%%%%%%%%%%%%%%%%%%%%%%%%%%%%%%%%%%%%%%%%%%%%%%%%%%%%%%%%%%%%%%%%%%%%%%%%%%%%

\subsection{The radial harmonic oscillator with broken SUSY}
The SUSY potential for the radial harmonic oscillator is given by
\begin{equation}
 \Phi(x)=x+\frac{\gamma}{x}\;.
\label{5.1}
\end{equation}
This SUSY potential leads to an unbroken SUSY system ($f=0$) if
the parameter $\gamma$ is negative. This case, which has already been
discussed in some detail in \cite{JuRo97}, leads to rather strict
conditions on the potential parameter $b$ and in turn gives rise to a very
limited class of new CES potentials. Therefore, we discuss here only the case
of broken SUSY, that is, $\gamma>0$.

The potential for the partner Hamiltonian $H_+$ reads
\begin{equation}
  \label{5.2}
 V_+(x)=\frac{x^2}{2}+\frac{\gamma(\gamma-1)}{2x^2}+\gamma+
\frac{b+1}{2}\;
\end{equation}
and gives rise to the following spectral properties
\begin{equation}
  \label{5.3}
E_n^+=2n+2\gamma+1+\frac{b}{2}\;,\quad
\psi_n^+(x)=
\left[\frac{2\,n!}{\Gamma(n+\gamma+1/2)}\right]^{1/2}x^{\gamma}\,
e^{-x^2/2}\,L_n^{(\gamma-1/2)}(x^2)\;.
\end{equation}
Positivity of $H_+$ leads us to the condition $b>-4\gamma -2$.

The general solution of eq.\ (\ref{3.7}) is expressed in terms of the
confluent hypergeometric function and reads
\begin{equation}
  \label{5.4}\textstyle
  u(x)=\alpha\, {}_1F_1(-\frac{b}{4},\gamma+\frac{1}{2},-x^2)
      +\beta\, x^{1-2\gamma}\,
       {}_1F_1(\frac{1}{2}-\frac{b}{4}-\gamma,\frac{3}{2}-\gamma,-x^2)\;.
\end{equation}
For small $0< x \ll 1$ this solution behaves like $u(x)\approx 
(\alpha + \beta x^{1-2\gamma})(1+O(x^2))$ and as a consequence we have to set
$\beta=0$ for SUSY to remain broken. Note that $\exp\{-\int
dx\,W(x)\}=\exp\{-x^2/2\}/x^\gamma u(x)$ and cf.\ eq.\ (\ref{3.11}).
Therefore, without loss of generality we set $\alpha=1$ and consider from now
on only the solution 
\begin{equation}
  \label{5.5}\textstyle
  u(x)={}_1F_1(-\frac{b}{4},\gamma+\frac{1}{2},-x^2)
      =e^{-x^2}{}_1F_1(\gamma +\frac{b+2}{4},\gamma+\frac{1}{2},x^2)
\end{equation}
leading to broken SUSY. This solution will have no zeros if $b>-4\gamma-2$, a
condition which we have found before from the positivity of $H_+$.

The partner potential reads
\begin{equation}
  \label{5.6}
  V_-(x)=\frac{x^2}{2}+\frac{\gamma(\gamma +1)}{2x^2}+\gamma-\frac{b+1}{2}+
\frac{u'(x)}{u(x)}\left(2x+\frac{2\gamma}{x}+\frac{u'(x)}{u(x)}\right)
\end{equation}
and is shown in Figure 5 for $\gamma=0.5$ and various values of $b$. As
expected there are singularities in $V_-$ for those values of $b$ which
violated the above condition. The eigenvalues of the corresponding Hamiltonian
$H_-$ are identical to those of $H_+$ given in (\ref{5.3}) with eigenfunctions
\begin{equation}
  \label{5.7}
  \psi_n^-(x)=
\left[\frac{2\,n!}{(n+\gamma+\frac{1}{2}+\frac{b}{4})
\Gamma(n+\gamma+1/2)}\right]^{1/2}
x^{\gamma+1}\,e^{-x^2/2}
\left(L_n^{(\gamma+1/2)}(x^2)+\frac{u'(x)}{2\,x\,u(x)}\right)
\end{equation}
which follow from the SUSY transformation (\ref{2.8}).

Finally, we note that for unbroken SUSY ($l=-\gamma > 0$) the special case
$b=0$ of (\ref{5.4})
\begin{equation}
  \label{5.8}
  u(x)=\alpha + 2\,\beta\,x^{2l+1} \int_0^x dt\, t^{2l}e^{-t^2}
\end{equation}
has, in essence,  been discussed before in \cite{Zhu87,AlFi88}.

%%%%%%%%%%%%%%%%%%%%%%%%%%%%%%%%%%%%%%%%%%%%%%%%%%%%%%%%%%%%%%%%%%%%%%%%%%%%%%

\subsection{The hydrogen atom }
The SUSY potential for the radial hydrogen atom problem is given by
\begin{equation}
\label{5.9}
\Phi(x)=\frac{a}{\gamma} - \frac{\gamma}{x}\;,\quad a,\gamma >0\;,
\end{equation}
and leads to the partner potential
\begin{equation}
  \label{5.10}
V_+(x)=-\frac{a}{x} +\frac{\gamma(\gamma +1)}{2x^2}
+\frac{1}{2}\left(b+a^2/\gamma^2\right)\;.
\end{equation}
The spectral properties of the associated partner Hamiltonian $H_+$ are well
known. For simplicity we give here only the discrete eigenvalues
\begin{equation}
  \label{5.11}
E_n^+=-\frac{a^2}{2(n+\gamma + 1)^2}+\frac{1}{2}\left(b+a^2/\gamma^2\right)\;,
\quad n\in{\mathbb N}_0\;.
\end{equation}
Then the positivity of $H_+$ leads to the condition
\begin{equation}
  \label{5.12}
  \rho = \sqrt{b+a^2/\gamma^2} > a /(\gamma +1)\;.
\end{equation}

In the present case the general solution of (\ref{3.7}) is again given in terms
of confluent hypergeometric functions
\begin{equation}
  \label{5.13}
  u(x)=e^{-(a/\gamma +  \rho)x}
\left[\alpha\,_{1}F_1(-\gamma-a/\rho,-2\gamma,2\rho x) +
\beta (2\rho x)^{2\gamma+1}\,_{1}F_1(\gamma+1-a/\rho,2\gamma+2,2\rho x)\right]
\end{equation}
and has the following asymptotic form for large $x$
\begin{equation}
  \label{5.14}
  u(x)=(2\rho x)^{\gamma-a/\rho}e^{(\rho - a/\gamma)x}
\left[\alpha \,\frac{\Gamma(-2\gamma)}{\Gamma(-\gamma-a/\rho)} +
\beta\,\frac{\Gamma(2\gamma+2)}{\Gamma(\gamma+1-a/\rho)}\right](1+O(x^{-1})).
\end{equation}

In order to find all conditions on the potential parameters we first note that
\begin{equation}
  \label{5.15}
  \psi_0^-(x)=\frac{C}{u(x)}\,x^\gamma e^{-ax/\gamma}
\end{equation}
and, therefore, the parameter
$\alpha$ must not vanish in order for SUSY to remain unbroken. Hence, without
loss of generality we may put $\alpha=1$. From the above asymptotic form
(\ref{5.14}) we can now also deduce further conditions on the parameters from
the positivity restriction on $u$. Summarizing all conditions we have 
\begin{equation}
  \label{5.16}
  \rho>\frac{a}{\gamma +1}\;,\quad
  \frac{\Gamma(-2\gamma)}{\Gamma(-\gamma-a/\rho)}>0\;,\quad
  \beta>-\frac{\Gamma(-2\gamma)}{\Gamma(-\gamma-a/\rho)}\,
        \frac{\Gamma(\gamma+1-a/\rho)}{\Gamma(2\gamma+2)}\;.
\end{equation}
In Figure 6 we give a graphical representation of the first and second
condition. Here the grey area shows the forbidden region due to the first
condition 
and the black area the forbidden region due to the second condition. In other
words, the allowed region of the two parameters $\gamma$ and $\rho$ for a
given coupling constant $a$, which is set equal to unity in Figure 6, is the
white area.

The CES potential for the hydrogen atom problem reads
\begin{equation}
  \label{5.17}
  V_-(x)=-\frac{a}{x} + \frac{\gamma(\gamma -1)}{2x^2}+ \frac{a^2}{\gamma^2} -
         \frac{\rho^2}{2}+\frac{u'(x)}{u(x)}
         \left(\frac{2a}{\gamma}-\frac{2\gamma}{x}+\frac{u'(x)}{u(x)}\right).
\end{equation}
Figure 7 shows this potential for $a=1$, $\beta=0$ and $\gamma
=2.8$. According to (\ref{5.16}) the allowed region for $\rho$ with the others
as fixed above is given by $]\frac{5}{16},\frac{5}{11}[\; \cup\;
]\frac{5}{6},5[$. 
The singularity appearing for $\rho\geq 5$ is clearly visible in Figure 7. The
other singularities are outside the plotted range of $0<x<2$ and therefore not
visible. In Figure 8 we keep $\gamma =2.8$, $a=1$ and $\rho = a/\gamma$ fixed
and show the potential (\ref{5.17}) for various values of $\beta$. Note the
singularity appearing for $\beta\leq -4.39554\times 10^{-4}$ according to the
violation of the last condition in (\ref{5.16}). Finally, let us also remark
that for the special case $\rho = a/\gamma$ (i.e.\ $b=0$) we have 
\begin{equation}
  \label{5.18}
  u(x)= \alpha + \beta (2\gamma +1)
  \int_0^{2a x/\gamma}dt\, t^{2\gamma}e^{-t}
\end{equation}
a result, which has previously been found in \cite{AlFi88}.

%%%%%%%%%%%%%%%%%%%%%%%%%%%%%%%%%%%%%%%%%%%%%%%%%%%%%%%%%%%%%%%%%%%%%%%%%%%%%%

\section{Quantum systems on a finite interval}
\setcounter{equation}{0}
As example for a quantum system defined on a finite interval we will consider
here the symmetric P\"oschl-Teller potential, whose SUSY potential is given by
\begin{equation}
  \label{6.1}
  \Phi(x)=\gamma\tan x\;,\quad \gamma >0\;,
\end{equation}
leading to a unbroken SUSY with
\begin{equation}
  \label{6.2}
  V_+(x)=\frac{\gamma(\gamma +1)}{2 \cos^2x}+\frac{b-\gamma^2}{2}\;.
\end{equation}
This is the well-studied P\"oschl-Teller potential, which gives rise to the
following spectral properties of $H_+$:
\begin{equation}
  \label{6.3}
  \begin{array}{l}
    \displaystyle
    E_n^+=\frac{1}{2}(\gamma + 1 + n)^2 +\frac{b-\gamma^2}{2}\;,\quad 
           n\in{\mathbb N}_0\;,\\[4mm]
    \displaystyle
    \psi_n^+(x)=\sqrt{\frac{(\gamma+1+n)\Gamma(2\gamma+2+n)}{\Gamma(n+1)}}
                \,\cos^{1/2}x \,P_{\gamma+n+1/2}^{-\gamma-1/2}(\sin x)\;.
  \end{array}
\end{equation}
Again, positivity leads to a condition on the parameter $b$,
$b>-2\gamma-1$. However, for later convenience we introduce another parameter
$\rho=\sqrt{\gamma^2 -b}$ and in terms of this, the above condition reads
\begin{equation}
  \label{6.4}
  0\leq \rho < \gamma+1\qquad\mbox{or}\qquad\rho\in i{\mathbb R}\;.
\end{equation}
The general solution for the corresponding differential equation (\ref{3.7})
is given in terms of hypergeometric functions
\begin{equation}
  \label{6.5}
  \textstyle
u(x)=\alpha\,{}_2F_1(-\frac{\gamma+\rho}{2},\frac{\gamma-\rho}{2},\frac{1}{2},
     \sin^2x) +
   \beta\sin x\,{}_2F_1(\frac{1-\gamma-\rho}{2},\frac{1-\gamma+\rho}{2},
      \frac{3}{2}, \sin^2x)\;.
\end{equation}
Obviously, as a necessary condition $\alpha$ must not vanish in order to have
no zeros in this solution. Hence, we will set $\alpha$ equal to unity in the
following discussion. From the values of $u$ at the boundaries of the
configuration space,
\begin{equation}
  \label{6.6}
  u(\pm\pi/2)=\frac{\Gamma(\frac{1}{2})\Gamma(\frac{1}{2}+\gamma)}
{\Gamma(\frac{1+\gamma+\rho}{2})\Gamma(\frac{1+\gamma-\rho}{2})}
\left[1\pm\frac{\beta}{2}\,
\frac{\Gamma(\frac{1+\gamma+\rho}{2})\Gamma(\frac{1+\gamma-\rho}{2})}
{\Gamma(1+\frac{\gamma+\rho}{2})\Gamma(1+\frac{\gamma-\rho}{2})}
\right],
\end{equation}
we also deduce a condition for the remaining parameter $\beta$:
\begin{equation}
  \label{6.7}
  |\beta|< 2
       \frac{\Gamma(1+\frac{\gamma+\rho}{2})\Gamma(1+\frac{\gamma-\rho}{2})} 
            {\Gamma(\frac{1+\gamma+\rho}{2})\Gamma(\frac{1+\gamma-\rho}{2})}\;.
\end{equation}

Finally, let us note that SUSY remains unbroken and the ground-state wave
function for $H_-$ is given by
\begin{equation}
  \label{6.8}
  \psi_0^-(x)=C\frac{\cos^\gamma x}{u(x)}\;.
\end{equation}
Hence, (\ref{6.4}) and (\ref{6.7}) constitutes the complete set of conditions
on the three parameters $\beta,\gamma$ and $\rho$.
The corresponding partner potential is given by
\begin{equation}
  \label{6.9}
  V_-(x)=\frac{\gamma(\gamma-1)}{2\cos^2x}-\gamma^2 +\frac{\rho^2}{2}+
 \frac{u'(x)}{u(x)}\left(2\gamma\tan x + \frac{u'(x)}{u(x)}\right)\;,
\end{equation}
which is shown in Figures 9-11 for some special cases. In Figure 9 and 10 we
have set $\beta =0$, $\gamma=2$ and chosen real $(0\leq \rho\leq 3.25)$ and
purely imaginary $(0\leq\rho/i\leq 4)$ values for $\rho$, respectively. Figure
9 exhibits singularities for $\rho\geq 3$ as expected from (\ref{6.4}),
whereas Figure 10 does not have singularities for the same reason. Finally,
Figure 11 shows the potential (\ref{6.9}) for fixed $\gamma=2$, $\rho=1$ and
various values of the asymmetry parameter $\beta$. Here due to condition
(\ref{6.7}) we expect and actually see singularities for $|\beta|\geq 2.35619$.

%%%%%%%%%%%%%%%%%%%%%%%%%%%%%%%%%%%%%%%%%%%%%%%%%%%%%%%%%%%%%%%%%%%%%%%%%%%%%%

\section{Concluding remarks}
\setcounter{equation}{0}
In this paper we have presented a method for constructing conditionally
exactly solvable potentials starting from the known SUSY potentials of
shape-invariant (exactly solvable) potentials. This method is more general then
those given in the literature before. In particular, most of the previously
constructed CES potentials correspond to the special case $b=0$ of our
method. 

There are several ways to generalize the present approach. Obviously, one
can now choose the newly found SUSY potentials of this paper as input and try
to construct further CES potentials from these. In general we expect to find a
hierarchy of new families of CES potentials belonging to the initial
shape-invariant one.  
In the present paper we have restricted ourselves to those parameter values
which conserve the nature of SUSY, that is, SUSY remains unbroken or broken by
adding the $f=u'/u$ term to the SUSY potential. This condition can
certainly be relaxed. Some of the conditions on the potential parameters
have been extracted from the asymptotic behaviour of the solution $u$ of
(\ref{3.7}). Hence, these conditions are only sufficient ones. In most cases
we expect them to be also necessary, but there may be exceptions. In any case,
if one wants to construct some exactly solvable model potential via the
present method a detailed analysis of the allowed parameter values is
advisable. 

We should also note that the present approach can be utilized to construct new
drift potentials for which the associated Fokker-Planck equation allows for an
explicit and exact solution. This would be similar to the discussion of the
linear harmonic oscillator by Hongler and Zheng \cite{HoZe82}. Let us also
mention that one may choose complex values for the parameters
$\alpha$ and/or $\beta$. This will lead to complex partner potentials $V_-$
whose associated non-hermitian Schr\"odinger Hamiltonian will have a real
spectrum \cite{BB98,ACDI97,CJT98}. Finally, we note that all the known 
shape-invariant potentials give rise to a dynamical group structure
\cite{BaRo92}.  This group structure induces, via the SUSY transformations
(\ref{2.7})-(\ref{2.8}), a related structure for the corresponding CES
Hamiltonian $H_-$. For example, one can construct from the well-known ladder
operators of the linear and radial harmonic oscillator via the supercharges
(\ref{2.3}) ladder operators for the corresponding partner Hamiltonian $H_-$.
It turns out that these operators close a non-linear algebra \cite{JuRo97}.
A detailed discussion, in particular, of the coherent states associated with
these non-linear algebras will be given elsewhere \cite{JuRo98}.

%%%%%%%%%%%%%%%%%%%%%%%%%%%%%%%%%%%%%%%%%%%%%%%%%%%%%%%%%%%%%%%%%%%%%%%%%%%%%%

%%%%%%%%%%%%%%%%%%%%%%%%%%%%%%%%%%%%%%%%%%%%%%%%%%%%%%%%%%%%%%%%%%%%%%%%%%%%%%
\newpage
\centerline{\bf Figure Captions}
\begin{itemize}
\item[Fig.\ 1:] The potential (\ref{4.3b}) for fixed $\alpha = 1$,
  $\beta=1.5\times\Gamma(\frac{b}{4}+1)/\Gamma(\frac{b+2}{4})$ and various
  ranges of the parameter $b$. Note that for $b\leq -2$ the potential exhibits
  singularities due to the existence of zeros in $u$ as given in (\ref{4.3}).
\item[Fig.\ 2:] The potential (\ref{4.3b}) for fixed $\alpha=1$, $b=-1.9$ and
  various values of the asymmetry parameter $\beta$. Here values of $\beta$
  with $|\beta|\geq 0.08569$ violate the positivity condition for $u$ (see
  text) and thus lead to singularities in $V_-$.
\item[Fig.\ 3:] The CES potentials of the Morse oscillator. Here $V_-$ is
  shown for $\alpha=0$, $\gamma=1$ and $\rho\in[0,4]$. The corresponding
  solution $u$ is given in (\ref{4.11}). Note the appearance of singularities
  in $V_-$ due to the violation of the conditions given in (\ref{4.12}).
\item[Fig.\ 4:] Same as Figure 4 but now for fixed $\gamma=\rho=3$ and various
  values of $\alpha$. Again singularities appear for $\alpha \leq -4/45 =
  -0.08889$ due to the last condition in (\ref{4.12}).
\item[Fig.\ 5:] The CES potential (\ref{5.6}) of the radial harmonic
  oscillator for $\gamma =0.5$ and various values of $b$. Note that the
  allowed range for this parameter is given by $b>-4\gamma-2=-4$.
\item[Fig.\ 6:] Allowed ranges for the parameters $\gamma$ and $\rho$ of the
  hydrogen atom problem according to the first two conditions given in
  (\ref{5.16}). For details see the text.
\item[Fig.\ 7:] The CES potential (\ref{5.17}) of the hydrogen atom problem
  for $a=1$, $\beta=0$, $\gamma = 2.8$ and various values of $\rho$. 
\item[Fig.\ 8:] Same as Figure 7, but for fixed $a=1$, $\gamma = 2.8$,
  $\rho=a/\gamma$ and various values of $\beta$.
\item[Fig.\ 9:] The CES potential (\ref{6.9}) of the P\"oschl-Teller problem
  for $\beta=0$, $\gamma = 2$ and $0\leq\rho\leq 3.25$.
\item[Fig.\ 10:] Same as Figure 9 but for complex $\rho$, $0\leq\rho/i\leq 4$.
\item[Fig.\ 11:] Same as Figure 9 with $\gamma=2$ and $\rho =1$ and various
  values of $\beta$.
%%%%%%%%%%%%%%%%%%%%%%%%%%%%%%%%%%%%%%%%%%%%%%%%%%%%%%%%%%%%%%%%%%%%%%%%%%%%%%
\end{itemize}

\newpage
\begin{thebibliography}{99}
\bibitem{Schr41}
E.\ Schr\"odinger, {\it Proc.\ Roy.\ Irish Acad.\ } {\bf 46A} (1940), 9.\\
E.\ Schr\"odinger, {\it Proc.\ Roy.\ Irish Acad.\ } {\bf 46A} (1941), 183.\\
E.\ Schr\"odinger, {\it Proc.\ Roy.\ Irish Acad.\ } {\bf 47A} (1941), 53.
\bibitem{InHu51}L.\ Infeld and T.E.\ Hull, {\it Rev.\ Mod.\ Phys.\ }
  {\bf 23} (1951), 28. 
\bibitem{Ju96}G.\ Junker, ``Supersymmetric Methods in Quantum and
  Statistical Physics,'' Springer-Verlag, Berlin, 1996.
\bibitem{Gen83}L.E.\ Gendenshtein, {\it JETP Lett.\ } {\bf 38} (1983),
  356.
\bibitem{AbMo80}P.B.\ Abraham and H.E.\ Moses, {\it Phys.\ Rev.\ A} {\bf 22}
  (1980), 1333.
\bibitem{Mie84}B.\ Mielnik, {\it J.\ Math.\ Phys.\ } {\bf 25} (1984), 3387.
\bibitem{Nie84}M.M.\ Nieto, {\it Phys.\ Lett.\ } {\bf 145B} (1984), 208.
\bibitem{LuPu86}M.\ Luban and D.L.\ Pursey, {\it Phys.\ Rev.\ D} {\bf 33}
  (1986), 431.\\
  D.L.\ Pursey, {\it Phys.\ Rev.\ D} {\bf 33} (1986), 1048.
\bibitem{Roy86}P.\ Roy and R.\ Roychoudhury, {\it Z.\ Phys.\ C} {\bf 31}
  (1986), 111.
\bibitem{Dar1882}G.\ Darboux, {\it Comptes Rendus Acad.\ Sci.\ (Paris)} {\bf
    94} (1882), 1456.
\bibitem{Dei78}P.A.\ Deift, {\it Duke Math.\ J.\ } {\bf 45} (1978), 267.
\bibitem{AnIoSp93}A.A.\ Andrianov, M.V.\ Ioffe and V.\ Spiridonov, {\it Phys.\
    Lett.\ A} {\bf 174} (1993), 273.\\
  A.A.\ Andrianov, M.V.\ Ioffe, F.\ Cannata
  and J.-P.\ Dedonder, {\it Int.\ J.\ Mod.\ Phys.\ A} {\bf 10} (1995), 2683.
\bibitem{BaSa95} V.G.\ Bagrov and B.F.\ Samsonov, {\it Teor.\ Mat.\ Fiz.} {\bf
    104} (1995), 356.\\
  V.G.\ Bagrov and B.F.\ Samsonov, {\it J.\ Phys.\ A} {\bf 29} (1996), 1011.
\bibitem{SoDu83}A.\ de Souza Dutra, {\it Phys.\ Rev.\ A} {\bf 47} (1993),
  R2435.\\
  R.\ Dutt, A.\ Khare and Y.P.\ Varshni, {\it J.\ Phys.\ A} {\bf 28}
  (1995), L107.
\bibitem{JuRo97}G.\ Junker and P.\ Roy, {\it Phys.\ Lett.\ A} {\bf 232}
  (1997), 155.\\
  G.\ Junker and P.\ Roy, ``Supersymmetric construction of exactly
  solvable potentials and non-linear algebras,'' preprint quant-ph/9709021, to
  appear in {\it Yad.\ Fiz.\ (Physics of Atomic Nuclei)}.
\bibitem{Wit81}E.\ Witten, {\it Nucl.\ Phys.\ } {\bf B188} (1981), 513;\\
   E.\ Witten, {\it Nucl.\ Phys.\ } {\bf B202} (1982), 253.
\bibitem{MaObSo66}W.\ Magnus, F.\ Oberhettinger and R.P.\ Soni, ``Formulas and
  Theorems for the Special Functions of Mathematical Physics,'' 3rd ed.,
  Springer-Verlag, Berlin, 1966.
\bibitem{HoZe82}M.-O.\ Hongler and W.M.\ Zheng, {\it J.\ Stat.\ Phys.} {\bf
    29} (1982), 317.\\
 M.-O.\ Hongler and W.M.\ Zheng, {\it J.\ Math.\ Phys.} {\bf 24} (1983), 336.
\bibitem{Dra83}P.G.\ Drazin, ``Solitons'', (Cambridge University Press,
  Cambridge, 1983).
\bibitem{Zhu87}D.\ Zhu, {\it J.\ Phys.\ A} {\bf 20} (1987), 4331.
\bibitem{AlFi88}N.A.\ Alves and E.D.\ Filho, {\it J.\ Phys.\ A} {\bf 21}
  (1988), 1011. 
\bibitem{BB98} C.M.\ Bender and S.\ Boettcher, {\it Real Spectra in
    Non-Hermitian Hamiltonians Having PT Symmetry}, preprint  physics/9712001,
  submitted to {\it Phys.\ Rev.\ Lett.}\\
 C.M.\ Bender and S.\ Boettcher, 
    {\it Quasi-exactly solvable quartic potential}, preprint physics/9801007,
  to appear in {\it J.\ Phys.\ A} (1998).
\bibitem{ACDI97}A.A.\ Andrianov, F.\ Cannata, J.-P.\ Dedonder and M.V.\ Ioffe,
  {\it SUSY quantum mechanics with complex superpotentials and real spectra}, 
  preprint (1997).
\bibitem{CJT98}F.\ Cannata, G.\ Junker and J.\ Trost, in preparation.
\bibitem{BaRo92}A.O.\ Barut and P.\ Roy, {\it in}
 ``Group Theory in Physics,'' AIP Conference Proceedings 266 
 (A.\ Frank, T.\ Seligman and K.B.\ Wolf eds.), p.\ 148, 
  American Institute of Physics, New York, 1992.
\bibitem{JuRo98}G.\ Junker and P.\ Roy, in preparation.
\end{thebibliography}
\end{document}